\RequirePackage[2020-02-02]{latexrelease}
\documentclass[prd, 10pt,aps,nofootinbib, floatfix, notitlepage,twocolumn,longbibliography,superscriptaddress]{revtex4-1}
\usepackage{placeins}
\usepackage{amsmath,amsfonts,amssymb}
\usepackage{mathrsfs}
\usepackage{graphicx}
\usepackage[english]{babel} 
\usepackage{ulem}
\usepackage{url} 
\usepackage{color}
\usepackage{bbold}
\usepackage{adjustbox}
\usepackage[utf8]{inputenc} 
\usepackage[colorlinks=true,citecolor=blue,urlcolor=blue]{hyperref}

\definecolor{green2}{RGB}{34,139,34}

\begin{document}
	
	\title{High Quality QCD Axion at Gravitational Wave Observatories}

	\author{Ricardo Z.~Ferreira}
	\email{rzambujal@ifae.es}
	\affiliation{Institut de F\'isica d'Altes Energies (IFAE) and The Barcelona Institute of Science and Technology (BIST), \\
		Campus UAB, 08193 Bellaterra, Barcelona, Spain
		\looseness=-1}
	\author{Alessio~Notari}
	\email{notari@fqa.ub.edu}
	\affiliation{Departament de F\'isica Qu\`antica i Astrofis\'ica \& Institut de Ci\`encies del Cosmos (ICCUB), Universitat de Barcelona, Mart\'i i Franqu\`es 1, 08028 Barcelona, Spain
		\looseness=-1}
	\author{Oriol~Pujolàs}
	\email{pujolas@ifae.es}
	\affiliation{Institut de F\'isica d'Altes Energies (IFAE) and The Barcelona Institute of Science and Technology (BIST), \\
		Campus UAB, 08193 Bellaterra, Barcelona, Spain
		\looseness=-1}
	\author{Fabrizio Rompineve}
	\email{fabrizio.rompineve@tufts.edu}
	\affiliation{Institute of Cosmology, Department of Physics and Astronomy, Tufts University, Medford, MA 02155, USA
		\looseness=-1}
	
	\date{\today}
	
	\begin{abstract}
		
		\noindent The axion solution to the strong CP problem is delicately sensitive to Peccei-Quinn breaking contributions that are misaligned with respect to QCD instantons. Heavy QCD axion models are appealing because they avoid this so-called ``quality problem''. We show that generic realizations of this framework can be probed by the LIGO-Virgo-KAGRA interferometers, through the stochastic gravitational wave (GW) signal sourced by the long-lived axionic string-domain wall network, and by upcoming measurements of the neutron and proton Electric Dipole Moments. Additionally, we provide predictions for searches at future GW observatories, which will further explore the parameter space of heavy QCD axion models.
		
	\end{abstract}

	\maketitle
	
{\bf Introduction}---A great amount of experimental effort has been aiming at discovering the QCD axion~\cite{Weinberg:1977ma, Wilczek:1977pj}, the pseudo-Goldstone boson of a spontaneously broken axial $U(1)$ Peccei-Quinn (PQ) symmetry~\cite{Peccei:1977hh, Peccei:1977ur} that explains the smallness of CP violation in strong interactions. 
 
	While attractive, the PQ mechanism is vulnerable to possible additional sources of symmetry breaking, generically misaligned with respect to the axion potential from QCD instantons. This {\it quality problem} (originally formulated with various perspectives in~\cite{Georgi:1981pu, Holdom:1982ex, Dine:1986bg, Holman:1992us, Kamionkowski:1992mf, Barr:1992qq, Ghigna:1992iv}) is alleviated in {\it heavy axion} models (see~\cite{Holdom:1982ex} and~\cite{Treiman:1978ge, Dimopoulos:1979pp, Tye:1981zy} for earlier related work), where a ``heavy QCD'' sector provides a larger contribution to the axion potential, aligned with that from QCD instantons.
	
	Existing realizations of this idea rely on: the QCD coupling becoming strong at high energies~\cite{Holdom:1982ex, Holdom:1985vx, Flynn:1987rs, Choi:1998ep, Agrawal:2017ksf, Kitano:2021fdl}, see also~\cite{Gherghetta:2020keg} for a 5D model; a separate confining gauge group, whose alignment is ensured by unification at high scales~\cite{Tye:1981zy, Rubakov:1997vp, Gherghetta:2016fhp, Gherghetta:2020ofz} or by a softly-broken $\mathbb{Z}_2$ symmetry~\cite{Berezhiani:2000gh, Dimopoulos:2016lvn, Hook:2019qoh}. When the strong coupling scale $\Lambda_{\text{H}}$ of the heavy sector is above the QCD scale $\Lambda_{\text{QCD}}$, the axion mass is larger than in the standard window, and the cosmological evolution of the axion field in the early Universe is shifted to higher energy scales. Despite its appeal, it is not immediately clear what the signatures of such a scenario are since generically the axion can be very heavy, e.g. above the electroweak scale, while its interactions remain very weak.\footnote{Axion masses and decay constants around or below the TeV scale can be probed at colliders, see e.g.~\cite{Bauer:2018uxu, Hook:2019qoh, Chakraborty:2021wda}.} Furthermore, in contrast to the standard case, a heavy QCD axion can easily decay in the early Universe, and thus leaves no detectable relic dark matter today.
	
	Nonetheless, in this Letter we show that heavy QCD axion models can be observationally probed at gravitational wave (GW) observatories (already at the currently operating LIGO-Virgo-KAGRA (LVK)~\cite{LIGOScientific:2014pky, VIRGO:2014yos, KAGRA:2018plz} interferometers), with the exciting possibility of a correlated signature in upcoming neutron and proton Electric Dipole Moment (nEDM, pEDM) measurements~\cite{Abel:2018yeo, Filippone:2018vxf}. 
	
	GWs are indeed radiated~\cite{Martin:1996ea} by the network of axionic topological defects (domain walls attached to strings)~\cite{Vilenkin:1982ks} (see also~\cite{Sikivie:1982qv}), which are abundant in the early Universe if the PQ symmetry is broken after inflation. In standard QCD axion models the network necessarily annihilates while making up only a very tiny fraction of the energy density of the Universe, and therefore the GW signal is too weak to be detectable~\cite{Hiramatsu:2012sc}. In contrast, the heavy QCD axion network can carry much more energy because of its larger domain wall tension. Furthermore, in generic realizations (e.g. DFSZ~\cite{Dine:1981rt, Zhitnitsky:1980tq} and simple generalizations of KSVZ models~\cite{Kim:1979if, Shifman:1979if}), the network can be long-lived while still avoiding the overproduction of relics, since radiated axion quanta are unstable. Annihilation of the network can be triggered by the misaligned PQ breaking effects that motivate the scenario in the first place (see~\cite{Holdom:1982ew}). These also induce a small but potentially observable shift of the QCD vacuum angle.

Our work points out a new source of observable gravitational waves from the dynamics of the QCD axion (see e.g.~\cite{VonHarling:2019rgb, DelleRose:2019pgi, Ramberg:2019dgi, Ramberg:2020oct} for previous work, unrelated to the axion quality problem and~\cite{Daido:2015gqa, Higaki:2016jjh, Chiang:2020aui, Gelmini:2021yzu} for related scenarios with ALPs).

{\bf The Heavy QCD Axion}---Heavy QCD axion models are characterized by an extra contribution to the axion potential that is larger than and aligned with the contribution from QCD. The zero temperature potential is %, $T\ll \Lambda_{\text{QCD}}$, 
	\begin{equation}
		\label{eq:Hpot}
		%V_a = \left[m_{a,\text{H}}^2+ m_{a}^2\right]f^2\left(1-\cos \frac{a}{f}\right),
		V_a = \left( \kappa^2_{\text{QCD}} \;\Lambda^4_{\text{QCD}}+\kappa^2_{\text{H}} \;\Lambda^4_{\text{H}} \right)\,\left(1-\cos \frac{a}{f}\right),
	\end{equation}
	where $f$ is the axion decay constant, $\Lambda_{\text{QCD,\,H}}$ denote the strong coupling scale of QCD and the heavy sector respectively, and $\kappa_{\text{QCD,\,H}} \leq1$ are prefactors that depend on details such as the fermionic spectrum. For instance, QCD gives $\kappa_{\text{QCD}}\simeq (m_u/\Lambda_{\text{QCD}})^{1/2}$ with $m_u$ the up quark mass. In explicit realizations of such scenarios \cite{Holdom:1982ex, Holdom:1985vx, Flynn:1987rs, Choi:1998ep, Agrawal:2017ksf, Gherghetta:2020keg,Tye:1981zy, Rubakov:1997vp, Gherghetta:2016fhp, Gherghetta:2020ofz,Berezhiani:2000gh, Dimopoulos:2016lvn, Hook:2019qoh} $\kappa_{\text{H}}\ll 1$ can similarly arise by the presence of a light quark in the heavy sector. Having QCD subdominant, the axion mass is dictated by the heavy sector as
	\begin{align}
		\label{eq:Hmass}
		m_{a} \simeq 10^8~\text{GeV}\left(\frac{10^{12}~\text{GeV}}{f}\right)\left(\frac{\Lambda_{\text{H}}}{10^{10}~\text{GeV}}\right)^{2}\kappa_{\text{H}}~.
	\end{align}

	For our discussion, it is important to recall that gauge instantons generically break the original $U(1)$ PQ symmetry to a discrete $\mathbb{Z}_{N_{\text{DW}}}$ subgroup, where $N_{\text{DW}}$ is a model-dependent integer number related to the axion coupling to gluons. Therefore the periodicity $2\pi f$ induced by the potential \eqref{eq:Hpot} can be smaller than the fundamental axion field range $2\pi f N_\text{DW}$ and the potential $V_a$ can feature $N_{\text{DW}}$ degenerate minima. In writing \eqref{eq:Hpot}, we assumed that the periodicity induced by the heavy sector coincides with that of QCD instantons. This appears to be linked to the requirement of alignment between the two sectors, as is evidently the case in constructions with a $\mathbb{Z}_2$ symmetry \cite{Hook:2019qoh} and in simple unification frameworks where SM and heavy sector fermions descend from the same fundamental representation of a higher-rank gauge group~\cite{Gherghetta:2016fhp,Agrawal:2017ksf}. This feature implies that the low-energy QCD-induced potential does not lift the degeneracy of the $N_{\text{DW}}$ minima.
	
Generically, however, we may expect further contributions to the axion potential, misaligned with $V_a$. Independently of its specific origin, such a contribution can be written as
	\begin{equation}
		\label{eq:bias}
		V_{b} \simeq - \mu_b^4 \cos \left(\frac{N_{\text{b}}}{N_{\text{DW}}}\frac{a}{f} - \delta \right),
	\end{equation}
	where $N_b$ defines the subgroup $\mathbb{Z}_{N_b}$ of the PQ symmetry which is preserved by \eqref{eq:bias} and $\delta$ is a CP violating phase. In the absence of tuning, this offset is naturally $O(1)$ and $\mu_b\ll \Lambda_{\text{H}}$ is required to solve the strong CP problem. The low temperature potential is $V=V_a+V_b$ and when $N_b=1$ or is co-prime with $N_{\text{DW}}$, the degeneracy of the $N_{\text{DW}}$ minima is lifted. In particular, the vacuum energy difference between the global CP preserving minimum and its nearest neighbor is of the order $\Delta V\simeq \mu_b^4[1-\cos(2\pi N_b/N_{\text{DW}})]$ (provided that $\delta$ is not too close to $\pi/N_{\text{DW}}$). Broadly speaking, \eqref{eq:bias} can originate at a scale $\Lambda_b$, such that $\mu_b = \kappa_b^{1/2} \Lambda_b$. $\Lambda_b \gg f\gg \Lambda_{\text{H}}$ can arise from UV physics via: non-perturbative effects, $\kappa_{b}\sim e^{-S/2}$ (see e.g. \cite{Dvali:2005an, Svrcek:2006yi, Hebecker:2018ofv}); higher-dimensional operators when the axion is the phase of a complex scalar field (see e.g.~\cite{Kamionkowski:1992mf}); another gauge sector with confinement scale $\Lambda_b$ and a light fermion of mass $m_q$, $\kappa_b\sim (m_q/\Lambda_b)^{1/2}$. $\Lambda_b \ll \Lambda_{\text{H}}$ can also arise from a confining gauge sector. Further details are provided in Appendix~\ref{sec:appA}.
	
Despite its smallness, a contribution from \eqref{eq:bias} can lead to potentially observable CP violation. In particular, at low temperatures one finds:
	\begin{equation}
		\label{eq:theta}
		\Delta\theta\equiv \theta-\theta_{\text{QCD}}\simeq r^4\left(\frac{N_b}{N_{\text{DW}}}\right)\left(\frac{\sin\delta}{\kappa_{\text{H}}^2}\right),
	\end{equation}
	where $r\equiv \mu_b/\Lambda_{\text{H}}$. Current bounds from nEDM measurements \cite{Abel:2020gbr} require $\Delta\theta\lesssim 10^{-10}$. Clearly, \eqref{eq:theta} shows that $\Lambda_{\text{H}}\gg \Lambda_{\text{QCD}}$ makes the PQ mechanism more robust against misaligned contributions.

	In the early Universe, the mass $m_a$ and the scale $\mu_b$ are generally temperature-dependent, for instance in the standard QCD axion case $m_a(T)\simeq m_{a} (T_0/T)^4$ for $T\geq T_0
	\simeq 134~\text{MeV}$ and $m_a(T)=m_a$ otherwise ~\cite{Borsanyi:2016ksw}. Nonetheless, our results are mostly independent of the detailed temperature dependence. 
	
{\bf Axionic Defects}---Let us now move to the cosmological evolution of topological defects, whose history begins at the PQ symmetry breaking scale $\sim N_{\text{DW}}f$. Our investigation concerns scenarios where this occurs during radiation domination after inflation, which will be generic for the values of $f$ considered in this work. 
	
	Axionic strings form at $T\lesssim N_{\text{DW}}f$ and continuously radiate axion quanta and gravitational waves. In the absence of significant friction due to the plasma, they quickly achieve a scaling regime~\cite{Kibble:1976sj, Kibble:1980mv} (see also \cite{Vilenkin:2000jqa} and~\cite{Gorghetto:2018myk, Hindmarsh:2019csc, Gorghetto:2020qws, Hindmarsh:2021vih} for recent updates), with energy density scaling as $\rho_{s}= \lambda \mu H^2$, with $\lambda$ a $O(1)$ parameter and $\mu\sim N_{\text{DW}}^2 f^2$ the string tension.
	
	This behavior is altered once $3H\simeq m_{a}(T)$. This occurs at a temperature $T_{\text{osc}}\gtrsim \Lambda_{\text{H}}$ (see Appendix~\ref{sec:appB}) when the axion field, with average initial value $a_i/ (N_\text{DW}f) \sim O(1)$, starts oscillating in its potential $V_a$ and domain walls form, attached to the strings, with a tension $\sigma\simeq 8 m_a(T) f^2$. At this epoch, two possibilities arise: i) when $N_{\text{DW}}=1$, the network of topological defects is rapidly annihilated by string-wall interactions (see e.g.~\cite{Vilenkin:2000jqa}); ii) when $N_{\text{DW}}>1$, the network persists because multiple domain walls pull each string in different directions.

	In both cases, the tension of the walls is larger than in the standard QCD axion case by a factor $\Lambda_{\text{H}}/\Lambda_{\text{QCD}}\gg 1$. For $N_{\text{DW}}>1$, in the absence of significant friction from the plasma (we show in Appendix~\ref{sec:appD} that this has a minor impact on our conclusions), the network rapidly achieves a scaling regime, with its energy density dominated by domain walls, $\rho_{\text{dw}}\simeq c \sigma H$, where $c$ is a $O(1)$ numerical prefactor (in this regime $V_a$ is normally already temperature-independent). This scales slower than matter and radiation and thus the network is potentially dangerous for cosmology~\cite{Zeldovich:1974uw}. However, domain wall domination can be generically avoided in the heavy axion scenario, thanks to the misaligned potential contribution $V_b$.  
	The resulting vacuum pressure causes the contraction of the false vacuum regions and the collapse of the network~\cite{Sikivie:1982qv} at a temperature $T_{\text{ann}}$, which can be estimated by imposing $\rho_{\text{dw}}\simeq \Delta V$ and more precisely determined via numerical simulations~\cite{Kawasaki:2014sqa}. Here we focus on the case where $V_b$ is temperature-independent below $\Lambda_{\text{H}}$, as occurs generically when PQ breaking is due to physics above $\Lambda_{\text{H}}$, see Appendix~\ref{sec:appB} for the temperature-dependent case. To set ideas and simplify expressions, in the following we set $N_b=1, N_{\text{DW}}=6$ as example values, fix numerical prefactors according to the simulations of~\cite{Kawasaki:2014sqa} and also fix the number of (entropy) relativistic degrees of freedom at $T_{\text{ann}}$ to the SM value at high temperatures $( g_{*s,\text{ann}}) \,g_{*,\text{ann}}=106.75$ (see also Appendix~\ref{sec:appB}), although our results are only mildly affected by these precise choices. We then find
	\begin{eqnarray}
		\label{eq:Tann}
		%\nonumber 
		T_{\text{ann}}\simeq \frac{10^{7}\text{GeV}}{\sqrt{\kappa_{\text{H}}}} %\left(\frac{106.75}{g_{*,\text{ann}}}\right)^{\frac{1}{4}} 
		\sqrt{\frac{10^{12}\text{GeV}}{f} }
		%\left(\frac{10^{12}\text{GeV}}{f} \right)^\frac{1}{2}
		\left(\frac{\Lambda_{\text{H}}}{10^{10}\text{GeV}}\right)\left(\frac{r}{0.005}\right)^2,
	\end{eqnarray}
showing that for $r\ll 1$ network annihilation is significant delayed.

	At \eqref{eq:Tann} the network collapses and its energy density is transferred mostly to mildly relativistic axion quanta~(see e.g.~\cite{Vilenkin:2000jqa} and~\cite{Kawasaki:2014sqa}). In contrast to the standard QCD axion case, in the heavy axion scenario these relics can efficiently decay to SM gluons, above the QCD Phase Transition (PT) and to photons and/or fermions, above and below the QCD PT, depending on the specific axion model. Focusing on the decay to gluons, since $m_a\gg~\text{GeV}$ in most of the parameter space of interest~\cite{Aloni:2018vki}, we find that decay is efficient below the temperature
	\begin{align}
		\label{eq:tdec}
		T_{a\rightarrow gg} &\simeq 10^7\text{GeV} ~\alpha_{s} \left(\frac{\sqrt{\kappa_{\text{H}}}\,\Lambda_{\text{H}}}{10^{10}\text{GeV}}\right)^3 \left(\frac{10^{12}\text{GeV}}{f}\right)^{\frac{5}{2}}
	\end{align}
	obtained by setting $\Gamma_{a\rightarrow gg}\simeq H$ (see Appendix~\ref{sec:appB}). This temperature can be larger than $T_{\text{ann}}$ for $r\lesssim  0.001$ and/or $f\lesssim 10^{12}~\text{GeV}$. Therefore, axion relics from the network will in general decay immediately.
	
	\begin{figure}[t]
		\includegraphics[width=\columnwidth]{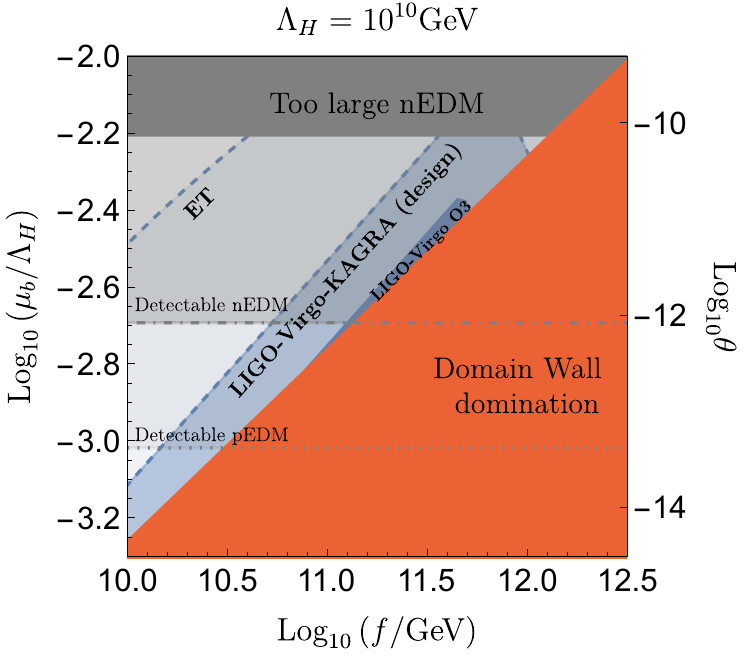}
		\caption{Regions of parameter space that can be probed by GW and/or nEDM experiments, for $\Lambda_{\text{H}}=10^{10}$ GeV and $\kappa_{\text{H}}=1$. Constraints are also shown, as dark-shaded regions, from: domain wall domination (lower right corner), nEDM \cite{Abel:2020gbr} (upper part), LIGO-Virgo O3 run \cite{Abbott:2021xxi} (dark-blue shaded). Dashed contours bound regions probed by LVK at design sensitivity and ET (sensitivity curves taken from~\cite{Schmitz:2020syl}). The gray shaded region will be also probed by neutron~\cite{Ahmed:2019dhe} (dot-dashed line) and proton EDM~\cite{Omarov:2020kws} measurements.}
		\label{fig:FixedLambda}
	\end{figure}
	
	Crucially, however, the string-wall network can source a  significant relic abundance of gravitational waves \cite{Vilenkin:1981zs, Preskill:1991kd, Chang:1998tb, Gleiser:1998na, Hiramatsu:2010yz, Kawasaki:2011vv, Hiramatsu:2013qaa}. The simple quadrupole estimate for their energy density $\rho_{\text{gw}}(T_\text{ann})\sim c^2\sigma^2/(32\pi M_p^2)$ has been confirmed by numerical simulations~\cite{Hiramatsu:2013qaa} (see also~\cite{Hiramatsu:2012sc}). Assuming a standard radiation-dominated cosmological history after domain wall annihilation, one finds that the relic abundance of gravitational waves today is 
	\begin{align}
		\label{eq:gws}
		\Omega_{\text{gw}}h^2&\simeq 0.01~(\Omega_{\text{rad}}^0h^2)~\tilde{\epsilon} \left(\frac{\rho_{\text{dw}}}{\rho_{\text{rad}}}\right)^2_{T=T_\text{ann}},
	\end{align}
	where $\tilde{\epsilon}\simeq 0.1-1$ is a numerical efficiency factor~\cite{Hiramatsu:2013qaa} and $\rho_{\text{rad}}$ and $\Omega_{\text{rad}}h^2\simeq 4\cdot 10^{-5}$ are the energy density and relic abundance of radiation today respectively. The formula above shows that when the network makes up $\gtrsim O(5\%)$ fraction of the energy density of the Universe at annihilation, its gravitational wave signal is detectable by present interferometers, i.e.~$\Omega_{\text{gw}}h^2\sim 10^{-9}$. This fraction at the annihilation temperature reads
	\begin{equation}
		\label{eq:fraction}
		\left.\frac{\rho_{\text{dw}}}{\rho_{\text{rad}}}\right\rvert_{T=T_{\text{ann}}}\simeq 0.1~\kappa_{\text{H}}^2\left(\frac{f}{10^{11}~\text{GeV}}\right)^2\left(\frac{0.003}{r}\right)^{4},
	\end{equation}
	for our example choice $N_b=1, N_{\text{DW}}=6$.
	
	The GW signal is peaked at a frequency corresponding to $H$ at annihilation (see e.g.~\cite{Hiramatsu:2012sc}). Redshifted to today:
	\begin{align}
		\label{eq:peakf}
		\omega_{\text{peak}}\simeq \frac{5~\text{Hz}}{\sqrt{\kappa_{\text{H}}}} \left(\frac{r}{0.005}\right)^2 \left(\frac{\Lambda_{\text{H}}}{10^{10}\text{GeV}}\right) \sqrt{\frac{10^{11}\text{GeV}}{f}} \,.
	\end{align}
	According to \eqref{eq:gws}, \eqref{eq:fraction} and \eqref{eq:peakf}, the signal from a heavy axion with $f\lesssim 10^{11}~\text{GeV}$, $\Lambda_{\text{H}} \gtrsim 10^{10}~\text{GeV}$ and $r \gtrsim 10^{-3}$ sits right in the reach of the LVK interferometers~\cite{LVdesign, LVKdesign}.
	
	The GW spectrum away from the peak frequency~\cite{Hiramatsu:2012sc} decreases as $\omega^{3}$ for $\omega<\omega_{\text{peak}}$, whereas for $\omega>\omega_{\text{peak}}$ it behaves as $\sim \omega^{-1}$, until a cutoff frequency corresponding to the domain wall width. However, further numerical simulations are required to understand the precise behavior of the spectrum around the peak frequency.

{\bf Predictions}---Although the $N_{\text{DW}}=1$ case does not leave observable GW signals (see Appendix~\ref{sec:appE}) due to the quick decay of the network, the situation is radically different for $N_{\text{DW}}>1$, where network annihilation is delayed. To simplify the presentation, we fix $N_b=1, N_{\text{DW}}=6, \kappa_{\text{H}}=1$ and $g_{*,\text{ann}}=g_{*s,\text{ann}}=106.75$ (see Appendix~\ref{sec:appB} for the case $\kappa_{\text{H}}\ll 1$) and $\delta=0.3$ and present results varying $\Lambda_{\text{H}}, r\equiv \mu_b/\Lambda_{\text{H}}$ and $f$. According to \eqref{eq:theta}, $r$ can then be traded for $\Delta\theta$.
	
	We first present results for large values of $\Lambda_{H}$, which maximally reduce the sensitivity to misaligned contributions.
	%\footnote{Reducing the sensitivity to higher dimensional operators with scale $\Lambda_b(f/\Lambda_b)^{\Delta/4}$ also requires $ f\ll \Lambda_b$~\cite{Kamionkowski:1992mf, Barr:1992qq, Ghigna:1992iv}.}

	We consider $V_b$ to be temperature-independent  at $T\lesssim \Lambda_{\text{H}} $ and $\Lambda_{\text{H}}$ to have a QCD-like temperature behavior (see also Appendix~\ref{sec:appC}).
	
	Fixing $\Lambda_{H}=10^{10}~\text{GeV}$ as a representative example, we show values of $r$ and $f$ that can be probed by gravitational wave observatories (dashed contours), together with constraints (solid contours), in Fig.~\ref{fig:FixedLambda}. In the lower right half, the string-wall network dominates before annihilation. While this region might not be completely ruled out, annihilation of the network in this case would require a dedicated study. In the upper part of the parameter space the PQ solution is spoiled, i.e. $\Delta\theta \gtrsim 10^{-10}$ \cite{Abel:2020gbr}. In the dark blue shaded region the GW signal is incompatible with the latest $2\sigma$ upper bound from LIGO-Virgo (LV)~\cite{Abbott:2021xxi}, this corresponds to the region close to DW domination. The dashed blue contours bound regions where the GW signal is detectable at the design sensitivity of LVK and Einstein Telescope (ET) respectively. The change of slope in the GW regions arises because of an intermediate phase of matter domination driven by the axions produced by the string-wall annihilation. This occurs at small decay temperatures~\eqref{eq:tdec}, corresponding to the right half of the figure. 
	
\begin{figure}[t]
		\includegraphics[width=0.95\columnwidth]{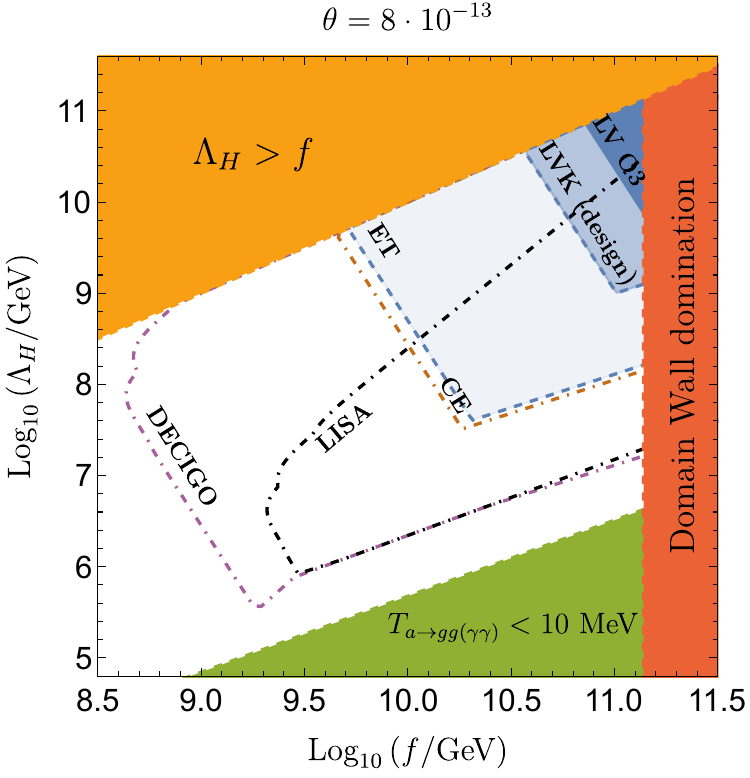}
		\caption{Regions of parameter space detectable at GW observatories, fixing $\theta=8\times 10^{-13}$ according to upcoming nEDM measurements~\cite{Ahmed:2019dhe} and $\kappa_{\text{H}}=1$. Same description and color code as in Fig.~\ref{fig:FixedLambda}, with the addition of CE, LISA and DECIGO's sensitivity curves (dot-dashed, taken from~\cite{Schmitz:2020syl}) and two constraints (dark-shaded regions) corresponding to axions decays below 10 MeV (lower right corner) and to $\Lambda_{\text{H}}>f$ (upper left corner).
	     }
		\label{fig:FixedTheta}
	\end{figure}

	Very interestingly, we find that a significant fraction of these GW-observable regions also predicts a detectable nEDM (and/or pEDM) in the near future~\cite{Abel:2018yeo, Ahmed:2019dhe, Omarov:2020kws}, i.e. $\Delta\theta\gtrsim 10^{-12}$ ($\Delta\theta\gtrsim 10^{-14}$) (above the dot-dashed and dotted gray contours respectively). Motivated by the exciting possibility of a combined heavy axion discovery via nEDM experiments and GW observatories, we fix $\Delta\theta=8\cdot 10^{-13}$ and broaden our analysis to different values of $\Lambda_{\text{H}}$ in Fig.~\ref{fig:FixedTheta}. We find that any $\Lambda_{\text{H}}\gtrsim 10^6~\text{GeV}$ leads to a GW signal in the foreseen reach of future ground (design LVK, ET, CE) and space based interferometers (LISA~\cite{2017arXiv170200786A, Baker:2019nia}, DECIGO~\cite{Yagi:2011wg, Isoyama:2018rjb}). The lower right corner in the figure is strongly constrained by the slow decay of axion quanta and a phase of matter domination which spoils BBN. As in Fig.~\ref{fig:FixedLambda}, the change of slope in the GW contours is due to an intermediate phase of matter domination, in the lower right part of the figure.
	
	Values of $\Lambda_{\text{H}}$ smaller than $10^6~\text{GeV}$ can also lead to viable cosmologies, observable GWs and detectable nEDM and/or pEDM, if the potential $V_b$ is temperature-dependent below $\Lambda_{\text{H}}$, see Appendix~\ref{sec:appC}.
	
\begin{figure}[b]
	\includegraphics[width=\columnwidth]{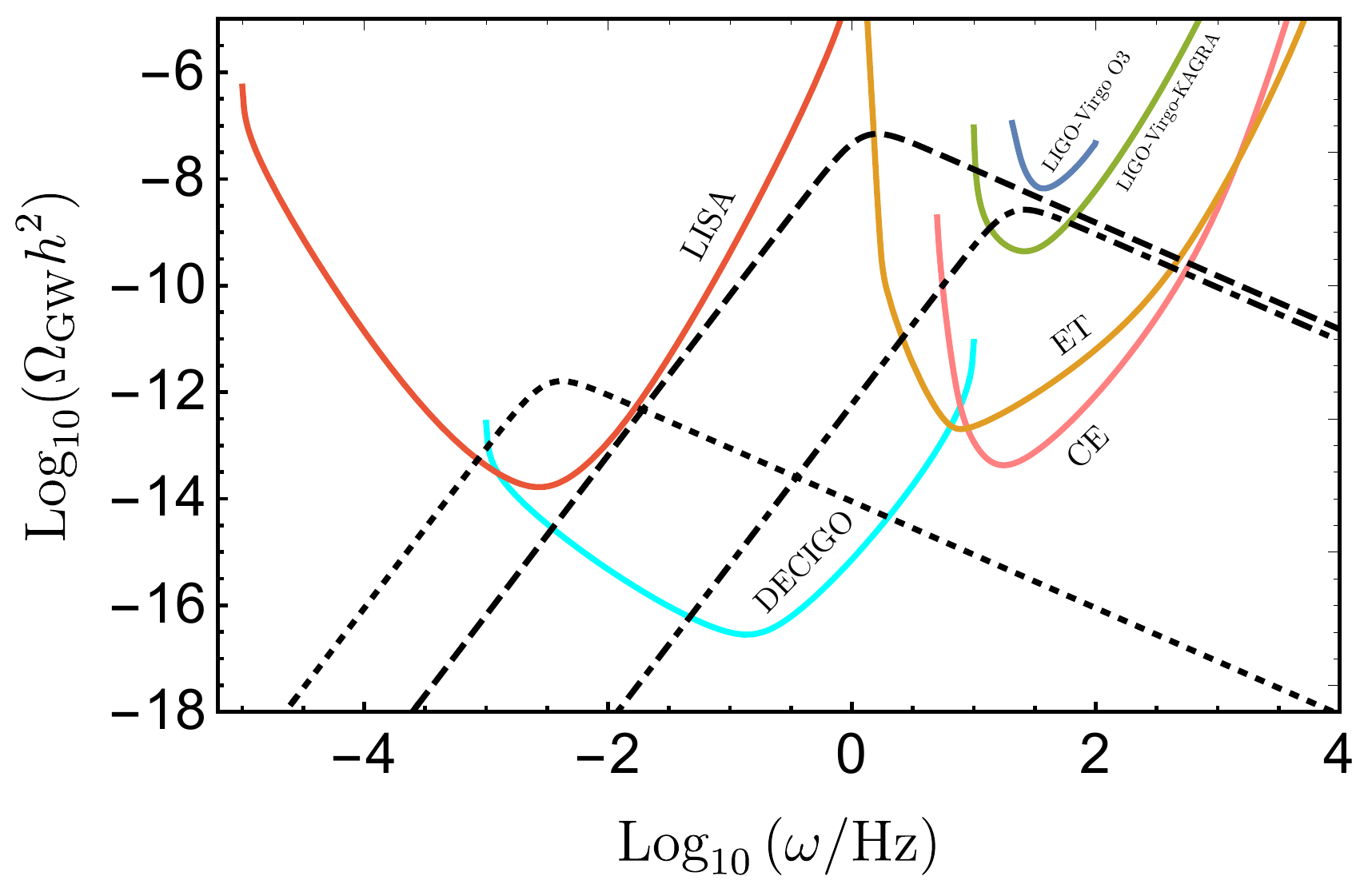}
	\caption{Representative GW spectra (dashed, dotted, dot-dashed lines) for $\kappa_{\text{H}}=1$ and $N_b=1, N_{\text{DW}}=6, \delta=0.3$. Dashed:  $\Lambda_{\text{H}}= 10^{10}~\text{GeV}$, $f\simeq 3\cdot 10^{11}~\text{GeV}$ and $\Delta \theta \simeq 9\cdot 10^{-11}$. Dotted: $\Lambda_{\text{H}}= 10^{7}~\text{GeV}$, $f=10^{10}~\text{GeV}$  $\Delta \theta \simeq 2 \cdot 10^{-12}$. Dot-dashed: $\Lambda_{\text{H}}= 10^{11}~\text{GeV}$, $f=1.6\cdot 10^{11}~\text{GeV}$ and $\Delta \theta \simeq 1.2 \cdot 10^{-11}$. Sensitivity curves are taken from~\cite{Schmitz:2020syl} and \cite{Abbott:2021xxi} for LIGO-Virgo O3. See also Appendix~\ref{sec:appB}.
	}
	\label{fig:full}
\end{figure}

	Finally, let us mention that for $\Lambda_{\text{H}}\lesssim 10^{10}~\text{GeV}$, LVK are expected to probe the high frequency tail of the GW signal ($\sim \omega^{-1}$), ET can investigate the peak and LISA can probe the low frequency tail ($\sim \omega^{3}$). Full GW spectra for some representative choices of parameters are shown in Fig.~\ref{fig:full}. 
	
A caveat is in order before our conclusions: the parameter space shown in Fig.~\ref{fig:FixedLambda} and Fig.~\ref{fig:FixedTheta} is further constrained if $V_b$ arises from dimension-five (and to a less relevant extent from dimension-six) operators with large coefficients. While we discuss this quantitatively in Appendix~\ref{sec:appA}, we note here that a large region of parameter space remains unaffected if such operators originate from non-perturbative effects (as can be expected if they are due to gravity, see e.g.~\cite{Dine:1986bg, Kallosh:1995hi, Svrcek:2006yi}).

{\bf Conclusions}---Heavy QCD axion models can feature a long-lived network of topological defects. The main finding of this Letter is that these models predict:  i) a stochastic gravitational wave signal measurable by the design LVK interferometers in a large region of parameter space (further broadened by ET and CE, with  the possibility of correlated signals also at LISA and DECIGO); ii) a nEDM and/or pEDM measurable in the near future; when: a) the new ``heavy QCD'' scale is large, i.e. $\Lambda_{\text{H}}\gtrsim 10^{10}~\text{GeV}$, thus making the PQ mechanism more robust; and b) misaligned PQ breaking terms that motivate these models in the first place are not strongly suppressed.
	
Furthermore, we showed that combined GW (at LISA and DECIGO) and nEDM/pEDM signatures also arise for $10^6~\text{GeV}\lesssim \Lambda_{\text{H}}\lesssim 10^{10}~\text{GeV}$.

	Our results do not strongly depend on the specific heavy QCD axion model, as long as its potential has approximately degenerate minima.
	
	We necessarily left several interesting points for future work. 
	First, in order to precisely characterize the GW signal, numerical simulations of axionic string-wall networks beyond the current literature, possibly including friction and plasma effects, are crucial~\cite{Kawasaki:2014sqa}.
	
	Second, we left unspecified the particle content and properties of the ``heavy QCD'' and of the misaligned PQ breaking sectors. However, these sectors may contain dark matter candidates (see e.g.~\cite{Garani:2021zrr}) or light states that can contribute to the number of extra relativistic degrees of freedom, $\Delta N_{\text{eff}}$~\cite{Aghanim:2018eyx, Abazajian:2016yjj}. 
	
	Furthermore, it is interesting to understand whether the collapse of the network of topological defects may also lead to a significant fraction of Primordial Black Holes, in a scaled-up version of the mechanism proposed in~\cite{Ferrer:2018uiu}.
	
	Finally, a more complete exploration of the parameter space of heavy axion models may lead to further interesting signatures. For instance, GWs of very low frequency may arise for misaligned sectors lighter than QCD, and may provide a new explanation for recent NANOGrav observations~\cite{Arzoumanian:2020vkk}. 
	An investigation of all these aspects is ongoing~\cite{Ferreira}.
	
	Our work is relevant for the ongoing effort to probe well-motivated regions in the parameter space of the PQ mechanism. Guided by the theoretical pursuit of ``higher quality'' models, we suggest that gravitational wave interferometers and nEDM/pEDM experiments may be the right laboratories to discover the heavy QCD axion.

{\bf Acknowledgments}---We thank Jaume Garriga, Alex Pomarol and Alex Vilenkin for useful discussions. The work of FR is supported in part by National Science Foundation Grant No. PHY-2013953. The work of RZF and OP was partly supported by the grants FPA2017-88915-P and SEV-2016-0588 from MINECO and 2017-SGR-1069 from DURSI. IFAE is partially funded by the CERCA program of the Generalitat de Catalunya.  The work of A.N. is supported by the grants FPA2016-76005-C2-2-P, PID2019-108122GBC32, ``Unit of Excellence Mar\'ia de Maeztu 2020-2023'' of ICCUB (CEX2019-000918-M), AGAUR2017-SGR-754.

\bibliography{biblio}
\bibliographystyle{BiblioStyle}

\appendix

\section{Origin of misaligned contributions}
\label{sec:appA}
	
\noindent In this section, we critically review well-motivated origins for misaligned contributions of the form 
\begin{equation}
V_{b} \simeq - \mu_b^4 \cos \left(\frac{N_{\text{b}}}{N_{\text{DW}}}\frac{a}{f} - \delta \right).
\end{equation} 
Such contributions can be generated at a scale $\Lambda_b$ such that $\mu_b = \kappa_b^{1/2} \Lambda_b$, with $\kappa_b\leq 1$.
	
Let us first consider misaligned terms generated at $\Lambda_b \gg f \gg \Lambda_{\text{H}}$. The following possibilities can then be envisioned:

\vspace{0.1cm}

\textbf{1 - Non-perturbative corrections to the axion potential}, among them: stringy and/or gravitational instantons, in which case $\kappa_{b}\sim e^{-S/2}$, where $S$ is the instanton action of interest (see e.g. the discussions in~\cite{Svrcek:2006yi} for stringy instantons and~\cite{Hebecker:2018ofv} for gravitational instantons); instantons of another gauge sector with confinement scale $\Lambda_b$ and a light fermion of mass $m_q$, giving $\kappa_b\sim (m_q/\Lambda_b)^{1/2}$.

\vspace{0.1cm}

\textbf{2 -  Higher-dimensional operators $\mathcal{O}_{\Delta}$} inducing 
		\begin{equation}
			V_{b}\sim c_{\Delta} \frac{\mathcal{O}_{\Delta}(\Phi)}{\Lambda_b^{\Delta-4}}+\text{h.c.},
		\end{equation} 
		with $\Delta>4$ in scenarios where the axion is the phase of a complex scalar field $\Phi$, whose renormalizable Lagrangian respects the PQ symmetry, giving $\kappa_{b}\sim \sqrt{c_{\Delta}}(N_{\text{DW}}f/\Lambda_b)^{\Delta/2}$. The size of the coefficients $c_{\Delta}$ depends on the specific origin of the higher dimensional operators.

\vspace{0.2cm}

The contributions above are often discussed as a source of the quality problem, since they induce the shift of the QCD vacuum angle given in eq. (4) of the main text and reported here for the reader's convenience: $\Delta\theta\simeq r^4 (N_b/N_{\text{DW}})(\sin\delta/\kappa_{\text{H}})$, with $r\equiv \mu_{b}/\Lambda_{\text{H}}$. Imposing that the QCD axion solution to the strong CP problem is kept, the following constraints on the two contributions above arise:

\vspace{0.1cm }
\textbf{1 -} For contributions with $\kappa_b\sim e^{-S/2}$, one finds:
\begin{align}
\label{eq:inst}
\nonumber S&\gtrsim 76+4\log\left(\frac{\Lambda_b}{10^{16}~\text{GeV}}\right)-4\log\left(\frac{\Lambda_{\text{H}}}{10^{10}~\text{GeV}}\right)\\
&-\log\left(\frac{\theta}{10^{-10}}\right)+\log\left(\frac{\sin\delta}{\kappa_{\text{H}}^2}\right),
\end{align}
whereas for contributions from confining gauge sectors with $\kappa_b\sim (m_q/\Lambda_b)^{1/2}$, one finds the following constraint on the light fermion mass:
\begin{equation}
\frac{m_{q}}{\Lambda_b}\lesssim 10^{-10}~\left(\frac{N_{\text{DW}}}{N_{b}}\right)\left(\frac{\kappa_{\text{H}}^2}{\sin\delta}\right)\left(\frac{\Lambda_{\text{H}}}{\Lambda_{b}}\right)^4\left(\frac{\theta}{10^{-10}}\right).
\end{equation}
In both cases, it is evident that taking $\Lambda_{\text{H}}$ as large as possible provides the most robust implementation of the PQ mechanism. In particular, the lower bound on the instanton action \eqref{eq:inst} is reduced by approximately a factor of two for $\Lambda_{\text{H}}\simeq 10^{10}~\text{GeV}$ with respect to the standard QCD axion. This can importantly relax the quality problem in certain setups~\cite{Dine:1986bg}.

\vspace{0.1cm}

\textbf{2 -} For contributions with $\kappa_b\sim  \sqrt{c_{\Delta}}(N_{\text{DW}}f/\Lambda_b)^{\Delta/2}$, one finds, focusing on the most important operators, i.e. those of dimension $\Delta = 5, 6$:
\begin{align}
\label{eq:dim5}
\nonumber \Lambda_{\text{H}}&\gtrsim 5~\text{TeV}\left(\frac{N_{\text{DW}}}{6}\right)\left(\frac{c_5}{\kappa_{\text{H}}^2}\right)^{\frac{1}{4}}\left(\frac{f}{10^{4}~\text{GeV}}\right)^{5/4}\\
&\times\left(\frac{10^{19}~\text{GeV}}{\Lambda_b}\right)^{\frac{1}{4}}\left(\frac{10^{-10}}{\theta}\right)^{\frac{1}{4}}, \quad \text{for}~\Delta=5,\\
\label{eq:dim6}
\nonumber \Lambda_{\text{H}}&\gtrsim 10^9~\text{GeV}\left(\frac{N_{\text{DW}}}{6}\right)^{\frac{5}{4}}\left(\frac{c_6}{\kappa_{\text{H}}^2}\right)^{\frac{1}{4}}\left(\frac{f}{10^{10}~\text{GeV}}\right)^{3/2}\\
&\times\left(\frac{10^{19}~\text{GeV}}{\Lambda_b}\right)^{\frac{1}{2}}\left(\frac{10^{-10}}{\theta}\right)^{\frac{1}{4}}, \quad \text{for}~\Delta=6.
\end{align}
The lower bounds above should be complemented with the upper bound $\Lambda_{\text{H}}\leq f$, which arises from consistency of the axion EFT. We have normalized $\Lambda_b$ to the Planck scale, to reflect the common assumption in the literature that the operators $\mathcal{O}_{\Delta}$ may be induced by gravitational interactions, see e.g.~\cite{Kamionkowski:1992mf}. 

The relevance of these constraints depends crucially on the size of the coefficients $c_{\Delta}$. It is sometimes assumed that $c_{\Delta}\sim O(1)$, which may be justified if UV physics, in particular gravity, breaks the PQ symmetry perturbatively. Under this assumption, \eqref{eq:dim5} imposes a very stringent constraint on $\Lambda_{\text{H}}$, which can be satisfied only for $5~\text{TeV}\lesssim \Lambda_{\text{H}}\lesssim f\sim 10~\text{TeV}$. This would severely limit the interesting region of the parameter space shown in Figs. 1 and 2 of our main text. 

\begin{figure*}[t]
		\includegraphics[width=0.9\columnwidth]{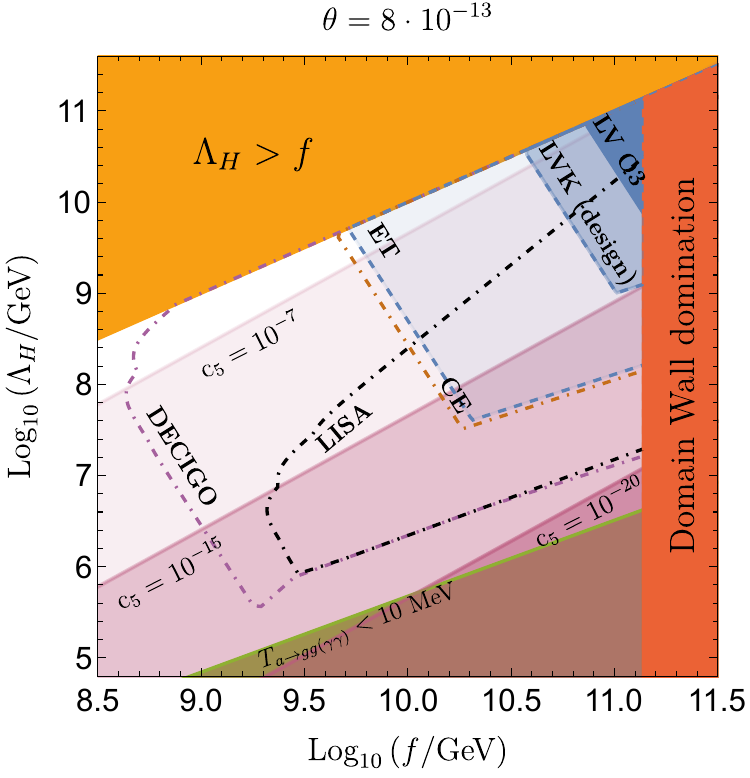}
		\includegraphics[width=0.9\columnwidth]{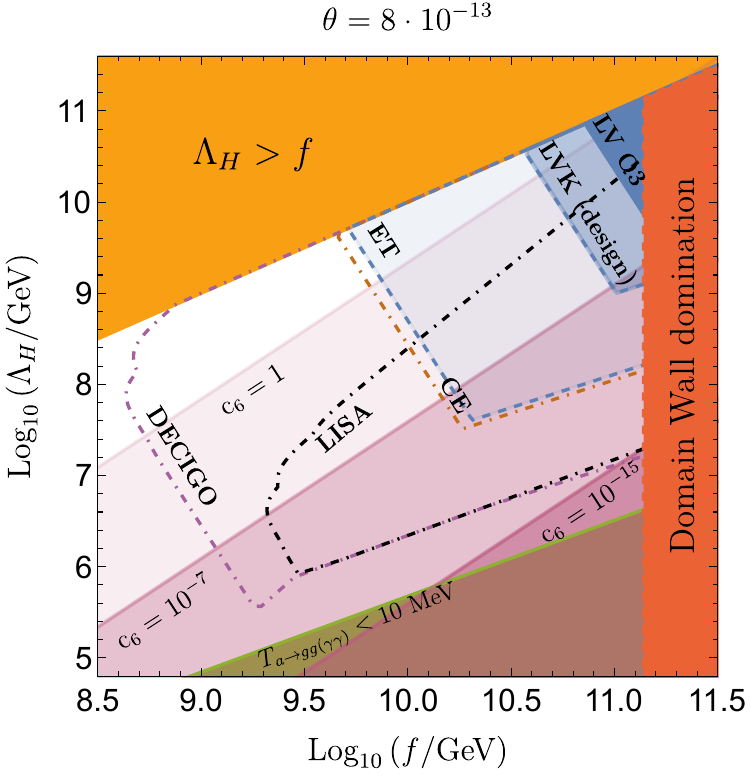}
	    \caption{{\it Left}: Same as Fig.~2 of the main text, with the constraint~\eqref{eq:dim5} for several choices of $c_{5}$ and with $\kappa_{\text{H}}=1$, under the assumption that Planck-suppressed dimension-5 operators which break the PQ symmetry are present. {\it Right}: same as the left figure, for dimension-6 operators.}
	    \label{fig:hd}
	\end{figure*}

On the other hand, arguments against global symmetries in gravitational theories suggest that the latter may break the PQ symmetry only via non-perturbative effects (see e.g.~\cite{Dine:1986bg,  Kallosh:1995hi, Banks:1996ea, Svrcek:2006yi}). In this case, higher dimensional operators may still be generated, albeit the coefficients $c_{\Delta}$ would then be expected to be exponentially suppressed by the corresponding instanton action, as in case $1$ above. Typical values $S_{\text{inst}}\sim O(100)$~(see e.g.~\cite{Svrcek:2006yi,Hui:2016ltb}) render Planck-suppressed operators irrelevant for the heavy QCD axion scenario considered in this work. Nonetheless, we may consider much smaller values of the action (as suggested by ) to understand what suppression would be required to affect our parameter space, as we do in Fig.~\ref{fig:hd} (left) for dimension $\Delta = 5$ operators and $\kappa_{\text{H}}=1$. We note that even for $c_{5}\lesssim 10^{-7}$, corresponding to a very small instanton action $S_{\text{inst}}\simeq 15$ (this roughly coincides with $S_{\text{inst}}\sim \ln M_p/f$ for $f\gtrsim 10^{11}~\text{GeV}$, which is the smallest estimate of the action from non-perturbative gravitational contributions~\cite{Kallosh:1995hi}), there exists a region of available parameter space where the GW signal is detectable at LIGO-Virgo-KAGRA (LVK), and that for $c_{5}\lesssim 10^{-15}$, corresponding to $S_{\text{inst}}\simeq 34$ the LVK region is entirely available.

It is also conceivable that dimension-5 operators may be forbidden in a concrete model. In this case, the most dangerous operators would be those of dimension 6. Constraints from~\eqref{eq:dim6} are shown in Fig.~\ref{fig:hd} (right). We note that in this case even the choice $c_{6}\sim 1$ leaves an available region of parameter space which overlaps with the LVK reach, and that for $c_{6}\sim 10^{-7}$ the entire region is available.

This is dramatically different from the situation in the standard (light) QCD axion case, where all operators up to dimension 9 (or 12 depending on the value of $f$) have to be forbidden in order for the axion to solve the strong CP problem.

\vspace{0.2cm}

Scenarios with $\Lambda_b \ll \Lambda_{\text{H}}$ can also arise from a confining gauge sector. Then $\kappa_b\sim (m_q/\Lambda_b)^{1/2}$ as in case $\mathbf{1}$ above. The quality problem in this case can then be evaded by taking $\Lambda_b$ sufficiently below $\Lambda_{\text{H}}$. Nonetheless, in contrast to the heavy QCD axion case, this is difficult in the standard QCD axion case where $\Lambda_b$ is bounded from below by the overproduction of axion dark matter from the decay of the axion string-wall network (see e.g.~\cite{Ferrer:2018uiu}).

\section{Network Evolution}
	\label{sec:appB}

Here we report full formulae for quantities defined in the main text, including the dependence on all parameters. We start with the temperature-dependent axion mass $m_{a,~\text{H}}$. We used the parametrization
	\begin{align}
	\label{eq:mass}
	\nonumber m_{a}(T)&=m_{a}\left(\frac{T_{0, H}}{T}\right)^{\gamma},\quad \text{for}~T\geq T_{0, \text{H}}\\
	m_{a}(T)&= \kappa_{\text{H}} \frac{\Lambda_{\text{H}}^2}{f},\quad \text{for}~T< T_{0, \text{H}}
	\end{align}
	where $T_{0, \text{H}}$ is a critical temperature, analogous to the critical temperature in QCD. Usually, $T_{0, \text{H}}\lesssim \Lambda_{\text{H}}$, for instance in QCD $T_{0}\simeq 134~\text{MeV}\simeq 0.4~\Lambda_{\text{QCD}}$~\cite{Borsanyi:2016ksw}. The exponent $\gamma$ can be estimated analytically at $T\gg \Lambda_{\text{H}}$ in the Dilute Instanton Gas Approximation (DIGA). For $G=SU(N_c)$ with $N_f$ light flavors, one finds~\cite{Gross:1980br}: $\gamma\simeq 11 N_c/6 + N_f/6 -2$.  Lattice calculations are required to compute $\gamma$ at $T\gtrsim \Lambda_{\text{H}}$. However, in the case of QCD, available lattice results agree well with the DIGA estimate~\cite{Borsanyi:2016ksw}. In order to produce the blue shaded region in our figures, we have fixed $\gamma=4, T_{0, \text{H}}=\Lambda_{H}$.
	
	The parameter $\kappa_{\text{H}}$ suppresses the heavy axion mass in models where $G$ contains at least one light fermion $\psi$, in which case $\kappa_{\text{H}} = \sqrt{m_{\psi}/\Lambda_{\text{H}}}$. Alternatively, $\kappa_{\text{H}}$ can also be much smaller than unity in models where $m_{a}$ arises from {\it small instantons} of QCD itself, when the latter becomes strongly coupled in the UV, above the EW phase transition~\cite{Holdom:1982ex, Holdom:1985vx, Flynn:1987rs, Choi:1998ep, Gherghetta:2020keg}. In this case SM quarks are massless and $\kappa_{\text{H}}$ is proportional to powers of their Yukawa couplings, see e.g.~\cite{Gherghetta:2020keg}.
	
	The temperature at which the heavy axion starts its oscillations is determined by imposing the condition $3H=m_a(T)$:
	\begin{align}
	\label{eq:tosc}
	\nonumber T_{\text{osc}}&\simeq  c_\gamma \cdot 10^{11}~\text{GeV}~\kappa_{\text{H}}^{\frac{1}{2+n}}\left(\frac{106.75}{g_{*}(T_{\text{osc}})}\right)^{\frac{1}{4+2n}}\\
	&\times\left(\frac{\Lambda}{10^{10}~\text{GeV}}\right)\left(\frac{10^{12}~\text{GeV}}{f}\right)^{\frac{1}{2+n}}.
	\end{align}
	where $c_\gamma$ varies by a $O(1)$ factor with $\gamma$. For $\gamma=4$ and $f=10^{12}~\text{GeV}$, one finds $T_{\text{osc}}\simeq 10^{11}~\text{GeV}$.
	For certain values of $f, r\equiv \mu_b/\Lambda_\text{H}$ and $\Lambda_{\text{H}}$, the heavy axion starts oscillating in the potential $V_b$ at temperatures above \eqref{eq:tosc}. Obviously, this can only happen if $V_b$ is larger than $V_a$ at high temperatures, for instance when $V_b$ is temperature-independent at temperatures slightly above $\Lambda_{\text{H}}$ (the temperature-dependent case is discussed in Appendix~\ref{sec:appC}). In this case, the temperature at which the axion would start oscillating around one of the minima of $V_b$ is given by
	\begin{align}
	\label{eq:Toscb}
	\nonumber T_{\text{osc, b}}&\simeq 5\cdot 10^9~\text{GeV} \left(\frac{\mu_b}{10^{7}~\text{GeV}}\right) \left(\frac{N_b}{N_\text{DW}}\right)^{\frac 1 2}\\
	&\times\left(\frac{106.75}{g_{*}(T_{\text{osc, b}})}\right)^{\frac{1}{4}}\left( \frac{10^{12}~\text{GeV}}{f} \right)^{\frac 1 2} .
	\end{align}
	In this case the scenario may still be cosmologically viable (if it satisfies other constraints), but the first domain walls to form are those of $V_b$, rather than of $V_a$. Their tension is significantly smaller than that of the $V_a$ walls, since $r\ll 1$. Furthermore, if $N_b=1$, then the $V_b$ network annihilates at high temperatures and there are no walls formed at $T_{\text{osc}}$. If $N_b$ is co-prime with $N_{\text{DW}}$, annihilation occurs close to $\Lambda_\text{H}$. Thus in this case the gravitational wave signal is suppressed compared to the case analyzed in the main text. However, we find that this occurs only in a tiny region of the parameter space, located in the upper left corner of Fig.~1 of the main text.
	
	When $T_{\text{osc, b}}\ll T_{\text{osc}}$, the network forms as described in the text. We assume that it quickly reaches the scaling regime, where $\rho_{\text{dw}}=c\sigma H(T)$. The annihilation temperature is then found by imposing the condition $d \sigma H = \Delta V$, where $d$ is a numerical parameter. One finds:
	\begin{align}
	\label{eq:Tannfull}
	\nonumber T_{\text{ann}}&\simeq 10^{8}~\text{GeV}~\frac{\sin(N_b\pi/N_{\text{DW}})}{\sqrt{c~d}\sqrt{\kappa_{\text{H}}}}\left(\frac{106.75}{g_{\text{ann}}}\right)^{\frac{1}{4}}\\
	&\times\left(\frac{10^{12}~\text{GeV}}{f}\right)^{\frac 1 2}\left(\frac{\Lambda_{\text{H}}}{10^{10}~\text{GeV}}\right)\left(\frac{r}{0.005}\right)^2,
	\end{align}
	where we kept the explicit dependence on all parameters, while in the text we fixed $N_b=1, N_{\text{DW}}=6, \kappa_{\text{H}}=1$ as representative example. The parameters $c$ and $d$ are determined by numerical simulations~\cite{Kawasaki:2014sqa} and depend on $N_{\text{DW}}$ (and presumably $N_b$). The relations to the parameters extracted from numerical simulations in~\cite{Kawasaki:2014sqa} are $c=2\mathcal{A}, d= 2 C_d$. For instance, for our example choice $N_{\text{DW}}=6$ the numerical simulations of~\cite{Kawasaki:2014sqa} find central values $\mathcal{A}=2.24, C_d= 3.14$ (we follow the $1\%$ criterion in~\cite{Kawasaki:2014sqa}), so that $c=4.48, d= 6.28$.
	
	The string-wall network may dominate the energy density of the Universe, if annihilation is too slow. The temperature corresponding to network domination is estimated by imposing $c\sigma H= 3H^2 M_{p}^2$. 
	\begin{align}
	\nonumber T_{\text{dw-dom}}&\simeq 6\cdot 10^6~\text{GeV} \sqrt{c\kappa_{\text{H}}} \left(\frac{106.75}{g_{*,\text{dw-dom}}}\right)^{\frac{1}{4}}\\
	&\times\left(\frac{f}{10^{12}~\text{GeV}} \right)^{\frac 1 2}\left(\frac{\Lambda_{\text{H}}}{10^{10}~\text{GeV}}\right).
	\end{align}
	In the shaded regions denoted with ``DW domination'' in Figures~1 and ~2 in the main text, $T_{\text{dw-dom}}>T_{\text{ann}}$. In this case it is in general problematic to gracefully exit the domain wall dominated phase, at least in the axion scenario considered in this work.
	
	The heavy axion decay rate into SM gluons (photons) is given by:
	\begin{align}
	\label{eq:rate}
	\Gamma_{a\rightarrow gg (\gamma\gamma)}=\frac{1}{64\pi}\left(\frac{c_{g(\gamma)}\alpha_{s (\text{em}})}{2\pi}\right)^2\frac{m_a^3}{f^2},
	\end{align}
	where $c_{g}=1$ and $c_{\gamma}=E/N_{\text{DW}}-1.92$, with $E$ being the model-dependent electromagnetic anomaly~(see e.g.~\cite{Zyla:2020zbs} for a review in standard DFSZ and KSVZ constructions) and the $-1.92$ arising only below the QCD PT. The decay is efficient once $\Gamma\simeq H$, which gives:
	\begin{align}
	\label{eq:tdecG}
	\nonumber T_{a\rightarrow gg (\gamma \gamma)}&\simeq 9 \cdot 10^6~\text{GeV}~\alpha_{s (\text{EM})}~\kappa_{\text{H}}^{3/2}\\
	 &\times\left(\frac{\Lambda_{\text{H}}}{10^{10}~\text{GeV}}\right)^3\left(\frac{10^{12}~\text{GeV}}{f}\right)^{\frac{5}{2}}
	\end{align}
	In models where $E/N\neq 0$, the decay channel to photons is in principle available above the EWPT, but it is less efficient than the channel to gluons (we use $\alpha_s=0.1$ in our Figures). Below the QCD PT, the latter remains available when $m_a\gtrsim 1~\text{GeV}$~\cite{Aloni:2018vki}, which happens in most of the parameter space of interest, otherwise the decay temperature should be computed according to the $a\rightarrow \gamma\gamma$ rate. In the shaded blue region denoted by ``Life-Time'' in Fig.~2 of the main text the decay temperature is below $10~\text{MeV}$. In this case products of axion decay can be dangerous for BBN. When the decay temperatures above are larger than the annihilation temperature of the network, the latter rapidly transfers its energy to the SM at $T_{\text{ann}}$.
	
	The annihilation of the string-wall network leads to a population of mildly relativistic axions, which soon contributes to the dark matter abundance. When the decay rates \eqref{eq:rate} are not fast enough, these axions can dominate the energy density of the Universe and lead to a temporary phase of matter domination. The temperature at which the matter dominated phase starts is found by imposing
	\begin{align}
	\nonumber &c\,\sigma H(T_{\text{ann}})\left(\frac{g_{*}(T)}{g_{*, \text{ann}}}\right)\left(\frac{T}{T_{\text{ann}}}\right)^3= 3 H(T) M_p^2 \nonumber
	\end{align}
	which gives 
	\begin{align}
	\nonumber T_{\text{MD}}= &8\cdot 10^6~\text{GeV}\left(\frac{\sqrt{d} (c\kappa_{\text{H}})^{3/2}}{\sin(N_b\pi/N_{\text{DW}})}\right)\left(\frac{106.75}{g_{*,\text{ann}}}\right)^{\frac{1}{4}}\\
	&\times\left(\frac{f}{10^{12}~\text{GeV}}\right)^{\frac{3}{2}}\left(\frac{\Lambda_{\text{H}}}{10^{10}~\text{GeV}}\right)\left(\frac{0.001}{r}\right)^2.
	\end{align}
	The temperature above should be compared with the decay temperatures \eqref{eq:tdecG}: whenever the former is larger, there is a phase of matter domination which should be taken into account when computing the gravitational wave signal. In principle, cold axions from the misalignment mechanism may also lead to an intermediate phase of matter domination. However, we find that in the regions of parameter space shown in the Figures, their contribution is subdominant compared to the axions from string-wall annihilation.

	Let us finally provide formulae for the gravitational wave signal. The peak frequency of the signal at the time of network annihilation is approximately given by the Hubble rate at that epoch, i.e. $\omega_{\text{peak}}\simeq H(T_{\text{ann}})$. Let us first consider the case in which the Universe is radiation dominated after domain wall annihilation. The redshifted frequency is then given by
	\begin{align}
\nonumber \omega_{\text{peak}}&\simeq 56~\text{Hz}~ \frac{\sin(N_b\pi/N_{\text{DW}})}{\sqrt{\kappa_{\text{H}} c~d}}\left(\frac{r}{0.005}\right)^2\\
  &\times\left(\frac{106.75}{g_{*, \text{ann}}}\right)^{\frac{1}{12}}\left(\frac{\Lambda_{\text{H}}}{10^{10}~\text{GeV}}\right)\left(\frac{10^{11}~\text{GeV}}{f} \right)^{\frac 1 2}.
	\end{align} 
	where, as mentioned in the main text, we have assumed $g_{*, \text{ann}}=g_{*s, \text{ann}}$ throughout the work.
	The amplitude of the signal at the peak frequency is instead given explicitly by~\cite{Hiramatsu:2013qaa}
	\begin{align}
	\label{eq:gwsfull}
	\nonumber\Omega_{\text{gw}}h^2&\simeq 0.01~(\Omega_{\gamma} h^2)~\tilde{\epsilon}\left(\frac{106.75}{g_{*,\text{ann}}}\right)^{\frac{1}{3}}\left(\frac{\rho_{\text{dw}}}{\rho_{\text{rel}}}\right)^2_{T=T_\text{ann}}\\
	\nonumber &\simeq 6\cdot 10^{-9}\frac{c^4 d^2  \tilde{\epsilon} \kappa_{\text{H}}^4}{\sin(N_b\pi/N_{\text{DW}})^4}\left(\frac{106.75}{g_{*,\text{ann}}}\right)^{\frac{1}{3}}\\
	&\times\left(\frac{f}{10^{12}~\text{GeV}}\right)^{4}\left(\frac{0.002}{r}\right)^{8},
	\end{align}
	where $\Omega_{\gamma}h^2\simeq 4\cdot 10^{-5}$ is the relic abundance of radiation today and $\tilde{\epsilon}$ is a numerical prefactor taken from simulations and of order $0.1-1$ \cite{Hiramatsu:2012sc}. To produce the plots we used $\tilde{\epsilon}=0.7$.
	
		\begin{figure}[t]
		\includegraphics[width=0.9\columnwidth]{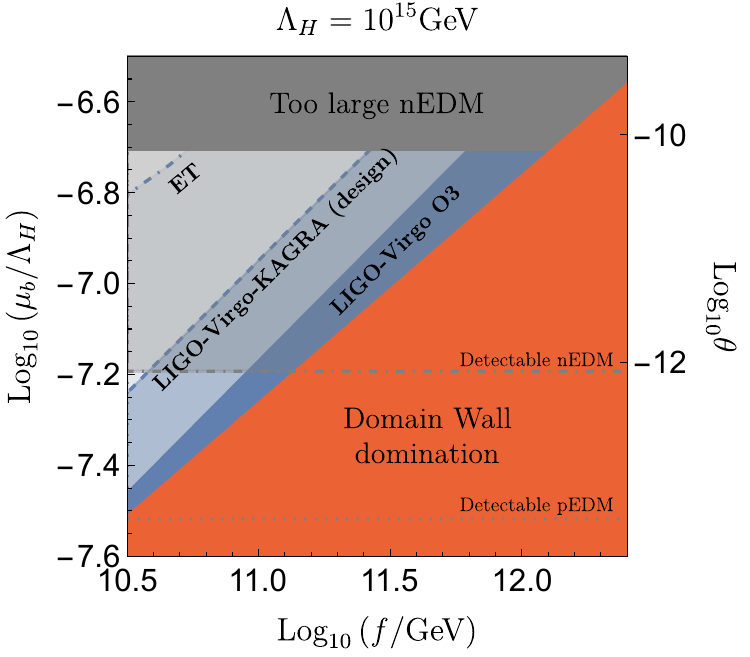}
		\caption{Same as Fig.~1 of the main text, but with $\Lambda_{\text{H}}=10^{15}~\text{GeV}$ and $\kappa_\text{H}=10^{-9}$.}
		\label{fig:smallkappafixedlambda}
	\end{figure}
	
	In the presence of an intermediate phase of matter domination, the formulae above should be adjusted to take into account the different redshift. In particular:
	\begin{align}
\nonumber\omega_{\text{peak, MD}}&\simeq H(T_{\text{ann}})\left(\frac{a(T_{\text{ann}})}{a(T_{\text{MD}})}\right)\\
&\times\left(\frac{H_{a\rightarrow gg(\gamma\gamma)}}{H_{\text{MD}}}\right)^{\frac{2}{3}}
	 \left(\frac{a(T_{a\rightarrow gg(\gamma\gamma)})}{a(\text{today})}\right),
	\end{align} 
	where we have used that $a\sim H^{-2/3}$ during matter domination. The Hubble rates at the beginning and end of the matter dominated phase can be determined by matching with the radiation dominated expression, i.e. $H\sim T^2/M_p$. One then gets:
	\begin{align}
	\omega_{\text{peak, MD}}\simeq \omega_{\text{peak}} \left(\frac{T_{a\rightarrow gg(\gamma\gamma)}}{T_{\text{MD}}}\right)^{\frac{1}{3}},
	\end{align} 
	which shows that the frequency can be much smaller than in the RD case, when $T_{\text{MD}}\gg T_{a\rightarrow gg(\gamma\gamma)}$. The amplitude of the signal is also suppressed by the duration of the MD epoch, i.e.
	\begin{equation}
	\nonumber \Omega_{\text{gw,MD}}h^2\simeq\Omega_{\text{gw}}h^2 \left(\frac{g_{*}(T_{a\rightarrow gg(\gamma\gamma)})}{g_{*}(T_{\text{MD}})}\right)^{\frac{1}{3}}\left(\frac{T_{a\rightarrow gg(\gamma\gamma)}}{T_{\text{MD}}}\right)^{\frac{4}{3}}.
	\end{equation}
	When the suppression of the amplitude due to MD is not too strong, the gravitational signal can be probed by interferometers which are sensitive at low frequencies, such as DECIGO, LISA and possibly PTAs.
	
	\begin{figure}[b]
		\includegraphics[width=0.85\columnwidth]{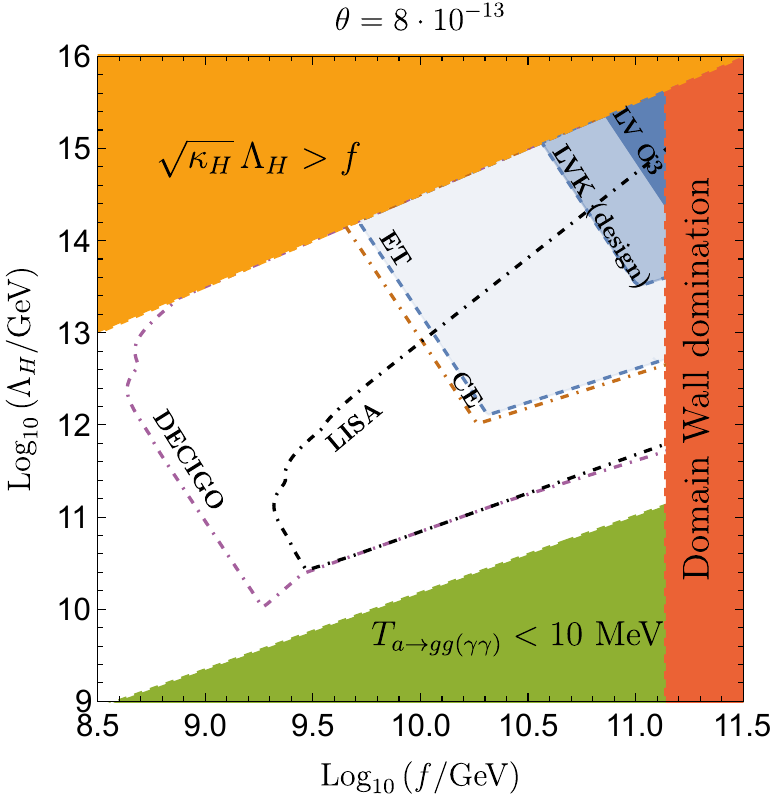}
	\caption{Same as Fig.~2 of the main text, but with $\kappa_\text{H}=10^{-9}$.}
		\label{fig:smallkappafixedtheta}
	\end{figure}

	The spectrum of GW waves is characterized by a peak at a frequency $\omega_\text{peak}$ roughly corresponding to the horizon at the time of the annihilation of the network, as explained in the main text. For smaller frequencies the spectrum decay as $\omega^3$, as dictated by causality, whereas for larger frequencies numerical simulations have shown a $1/\omega$ dependence which is cut off at a frequency corresponding to the axion mass \cite{Hiramatsu:2012sc}. The details of the spectrum near the peak are subject to more uncertainties and motivate further numerical studies. In this work we consider the peak to have a $O(1) \omega_\text{peak}$ width, as suggested by numerical simulations, and use the following simple formula for the spectrum to produce Fig.~3 of the main text:

	\begin{eqnarray}
	   \Omega_{\text{gw}} (\omega)= \Omega_{\text{gw}} (\omega_\text{peak}) \frac{4}{\left(\frac{\omega}{\omega_{\text{peak}}}\right)^{-3} + 3\left(\frac{\omega}{\omega_{\text{peak}}}\right)}
	\end{eqnarray}
	Smaller widths would not significantly change the predictions. If the peak width is much larger than $O(1) \omega_\text{peak}$ the detectability of the signal would significantly improve. 
	
	Additionally, we present results for a representative example with $\kappa_{\text{H}}\ll 1$ in Figures~\ref{fig:smallkappafixedlambda} and~\ref{fig:smallkappafixedtheta}. In this case, we take $\sqrt{\kappa_{\text{H}}}f\leq f$ as cutoff of the axion effective theory. Correspondingly, LIGO-Virgo-KAGRA can now probe much larger scales $\Lambda_{\text{H}}\lesssim 10^{15}~\text{GeV}$ than in the case $\kappa_{\text{H}}=1$.

	\section{Temperature dependent $V_b$}
	\label{sec:appC}

\begin{figure}[t]
		\includegraphics[width=0.9\columnwidth]{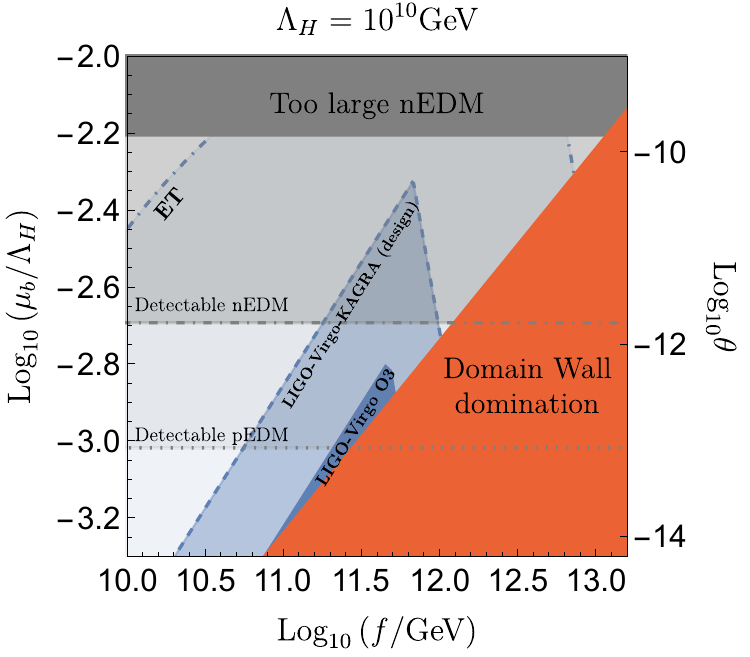}
		\caption{Same as Fig.~1 of the main text, but with a temperature dependent bias.}
		\label{fig:FixedLambda-Tdep}
	\end{figure}

\begin{figure}[t]
		\includegraphics[width=0.85\columnwidth]{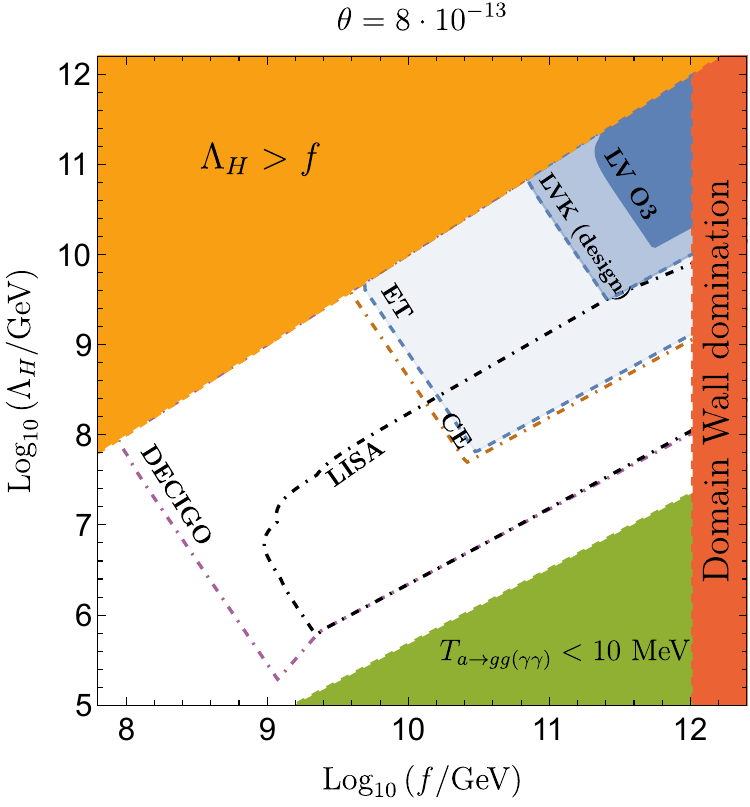}
		\caption{Same as Fig.~2 of the main text, but a with temperature dependent bias.}
		\label{fig:Fixedtheta-Tdep}
	\end{figure}

	In the main text we focused on the case where the misaligned contribution to the axion potential is temperature independent. This is typically the case when the source of PQ breaking comes from energy scales above $f$. Instead, if there are infrared contributions, e.g. coming from sectors which confine at scales $\mu_b \ll f$, the axion potential will be temperature dependent and saturate at temperatures $T_b \lesssim \mu_b$. In this Section we present the results for the latter case and consider a bias potential given by
	\begin{equation}
		\label{eq:biasB1}
		V_{b}(T) \simeq \mu_b^4(T) \cos \left(\frac{N_{\text{b}}}{N_{\text{DW}}}\frac{a}{f} + \delta \right),
	\end{equation}
	where $ \mu_b^4(T) = \mu_b^4 \min \left[\left(\frac{T_b}{T}\right)^n,1 \right] $.
	When $n=0$ we recover the case described in the main text. On the other hand, for large $n$ the potential is strongly temperature dependent, the potential approaches a step function and the predictions become almost independent of $n$. For simplicity, we consider the case of a temperature dependence similar to QCD, i.e,  $T_b \lesssim  \mu_b$ and $n=4$.
	The results are shown in figures \ref{fig:FixedLambda-Tdep} and \ref{fig:Fixedtheta-Tdep} for the same values used in the temperature independent case (see Figures~1 and~2 of the main text). 
	
	The main consequence of a temperature dependent bias is that the annihilation of the network will happen at a later temperature $T \lesssim T_b$ thus causing an enhancement of the relative energy in the network, $\rho_\text{dw}/\rho_\text{tot}$, and so on the GW signals. This allows to probe a larger region of parameter space, in particular smaller values of $f$, simultaneously with nEDM and GW experiments.

	\section{Friction}
	\label{sec:appD}
	
	Particles in the primordial plasma that interact with the domain wall can exert thermal friction \cite{Zeldovich:1974uw,Everett:1974bs,Kibble:1976sj}.
	 This is usually negligible for the QCD axion, since the axion mass is very small compared to the temperature of the plasma, but it can be relevant in the heavy axion scenario. Quantifying these effects is a model-dependent task, since axion interactions are not fixed by the requirement that the axion solves the strong CP problem. 
	 Here, we consider a conservative case, where we assume one particle in the plasma to have a probability of order one to be reflected by the axion wall, and show that even in this case friction is only important in a region of parameter space relevant for future interferometers. Even so, we find only a minor impact on the GW signal in this case.
	
	The thermal pressure exerted on the wall by relativistic particle in thermal equilibrium is given by \cite{Vilenkin:1981zs}
	\begin{eqnarray} \label{eq: Friction force}
		F=   P_r  \,n \, \Delta p \, 
	\end{eqnarray}
	where $n=g_\text{fr}\, \zeta(3) T^3/\pi^2$ is the density of relativistic particles that are reflected by the wall with a probability $P_r$, $g_\text{fr}$ is the number of degrees of freedom and $\Delta p = v T$ the momentum transfer (for non-relativistic walls and after averaging over the two sides of the wall \cite{Vilenkin:1981zs}).
	If the friction caused by the thermal pressure, $F/\sigma$, overcomes the Hubble friction, $H$, the wall will reach an attractor regime where the self-acceleration due to its own tension balances the friction force \cite{Vilenkin:1981zs}
	\begin{eqnarray} \label{eq: Balance of forces}
		\sigma/R \simeq F \,.
	\end{eqnarray} 
	where $R$ is the typical curvature of the network.
	In this attractor, the velocity of the wall increases with time as 
	\begin{eqnarray}
		v(t)\simeq 8.7 \left(\frac{1}{g_\text{fr} P_r } \right)^{1/2} \left(\frac{\sigma}{M_p T^2} \right)^{1/2}
	\end{eqnarray}
	where we used eq. \ref{eq: Friction force} and  $R=vt$. The friction domination regime ends when $F/\sigma \lesssim H$ or, equivalently, when the wall reaches the relativistic limit. Interestingly, when the background is radiation dominated, this always happen at a temperature $T_\text{fr}$ close to the time of domain wall domination $T_\text{fr} \simeq 0.43 \sqrt{g_\text{fr} \,P_r} \,T_\text{dw-dom}$.
	
	Another consequence of thermal friction is the delay of the network annihilation. In the absence of friction, the wall annihilates when the bias pressure overcomes the domain wall tension, $V_b \gtrsim \rho_\text{dw}$. However, because $R$ is now smaller due to friction, $\rho_\text{dw}$ is enhanced and so it will take more time for the bias pressure to dominate and cause the collapse of the network. For a temperature independent bias in a radiation dominated era, that happens when
	\begin{eqnarray}
		H_\text{ann}^\text{fr} = 1.2\, \left(\frac{g_{*,\text{ann}}}{g_\text{fr} \, P_r}\right)^{1/3}  H_\text{ann}^{2/3} \,H_\text{dom}^{1/3}  < H_\text{ann}
	\end{eqnarray}
	where $H_\text{ann},H_\text{dw-dom}$ are the Hubble rates at the time  of domain wall annihilation and domination, respectively, in the absence of friction.

	The small velocity of the domain walls also affects the spectrum of gravitational waves. Using the expression for the quadrupole formula one can estimate the power in gravitational waves at the time of annihilation as \cite{Nakayama:2016gxi}
	\begin{eqnarray}
	P^\text{fr}_{GW} = G \left(\frac{d^3 I}{dt^3 } \right)^2 \sim  G  \left(  v^3 \sigma R\right)_{\text{ann}}^2 
	\end{eqnarray}
	where we used $I \sim  \sigma R^4 $ for the quadrupole moment of the wall.
	There are three main changes compared to the frictionless case: i) the amount of gravitational waves is suppressed by $\left(v^3R H\right)^2$, ii) the network annihilates later thus increasing the relative energy in the GWs (when compared to the background), iii) the peak frequency of the spectrum at the time of annihilation is expected to move from $H^\text{fr}_\text{ann}$ to $1/R^\text{fr}_\text{ann}$. 
	
		\begin{figure}
		\centering
		\includegraphics[width=\columnwidth]{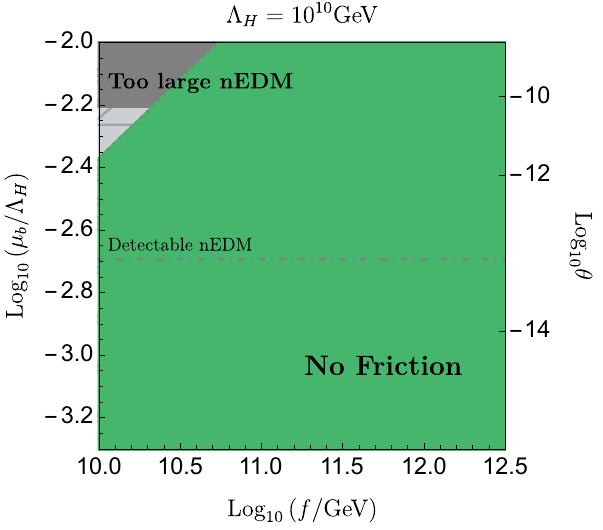}
		\caption{Regions of parameter space where thermal friction on the domain walls from gluons can be neglected, for $\Lambda=10^{10}~\text{GeV}$, using the same choices as for Fig.~1 of the main text.}
		\label{fig:lambdafixed-friction-labelsinside}
	\end{figure}
	
	What remains to assess is the value of the reflection probability $P_r$. To do that we need to specify the interactions of the axion with the plasma. At temperatures above $1$ GeV, the focus of this work, there is one unavoidable interaction due to the axion coupling to gluons. Other interactions are also possible, for example with SM fermions, which can be sizeable in DFSZ-like constructions, or with particles in the dark sector (both in the bias or in $\Lambda_H$), but those are model dependent.
	
\begin{figure}[t]
		\includegraphics[width=0.85\columnwidth]{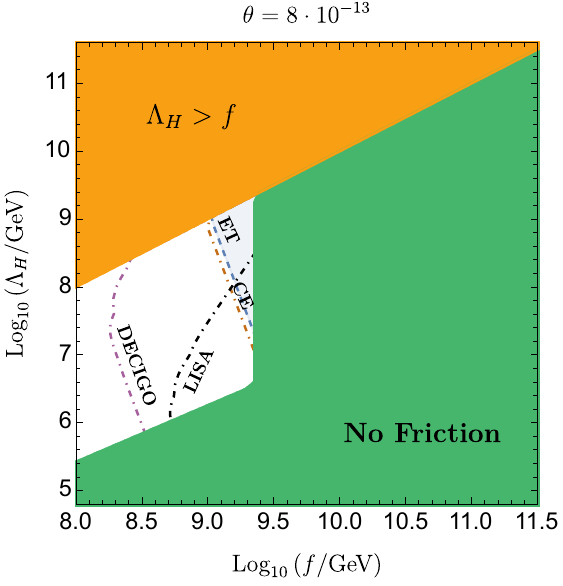}
		\caption{Regions of parameter space where thermal friction on the domain walls from gluons can be neglected, for $\theta=8\cdot 10^{-13}$, using the same choices as for Figure~2 of the main text. Regions that can be probed by GW interferometers are also shown in the regions where friction is important and is thus included in the calculation.}
		\label{fig:thetafixed-friction}
	\end{figure}

	Here we will take a conservative approach and assume the reflection probability to be maximal, $P_r=1$. We will present a more detailed analysis of friction due to gluons in \cite{Ferreira}. 
	In Figs. \ref{fig:lambdafixed-friction-labelsinside} and \ref{fig:thetafixed-friction} we show the region of parameter space where friction is important, assuming $g_\text{fr}=1$, and the respective prediction for the GW spectrum. In order to perform a fair comparison with the frictionless case, we also included the numerical factors $c,d,\beta$ in the expressions for friction. The results show that friction is only relevant in a small region of parameter space which is bounded by the condition that the network annihilates during the friction domination regime $T^\text{fr} < T_\text{ann}^\text{fr}$ and by the requirement that $T^\text{fr}_\text{ann}< m_a$, otherwise the reflection probability will be suppressed. Even in the corner where friction is relevant, the predictions are only mildly different from those without friction (c.f. Figs.1 and 2 of the main text).

	\section{$N_{\text{DW}}=1$}
	\label{sec:appE}
	
	Here we discuss the case $N_{\text{DW}}=1$ in some more detail. In contrast with the case  $N_{\text{DW}}>1$, the network is unstable even in the absence of extra sources of PQ breaking, see e.g.~\cite{Vilenkin:2000jqa} and~\cite{Hiramatsu:2012sc, Kawasaki:2014sqa} for numerical simulations. Annihilation occurs once the energy density in strings is comparable to that in domain walls. The energy density of strings is $\rho_s\simeq \lambda \mu H^2$, where $\mu=\pi f^2 \log(t~f/\sqrt{\lambda})$ is the string tension, $t$ is time, and $\lambda\simeq 4$ is a parameter that can be determined via numerical simulations. Deviations from this regime that have been observed in the literature \cite{Gorghetto:2018myk, Gorghetto:2020qws} are not particularly relevant to us, since in the heavy axion scenario strings only evolve until $\Lambda_{\text{H}}\gg \Lambda_{\text{QCD}}$.
	By equating to $\rho_{\text{dw}}=c \sigma(T) H$ (for $N_{\text{DW}}=1$, $c\approx 1$~\cite{Kawasaki:2014sqa}), one finds the annihilation temperature, which also depends on the temperature dependence of the heavy axion potential. We follow here the same example choice as in Appendix~\ref{sec:appB}, although this does not affect our conclusions.
	\begin{align}
	\nonumber T_{\text{ann}}&\simeq 10^{10}~\text{GeV}\left(\frac{2\kappa_{\text{H}}~c}{\lambda}\right)^{\frac{1}{6}}\left(\frac{106.75}{g_{*}}\right)^{\frac{1}{12}}\\
	&\times \left(\frac{10^{15}~\text{GeV}}{f}\right)^{\frac{1}{6}}\left(\frac{\Lambda_{\text{H}}}{10^{10}~\text{GeV}}\right).
	\end{align}
	Following the same steps as in Appendix~\ref{sec:appB}, and assuming radiation domination below $T_{\text{ann}}$, the gravitational wave peak frequency is found to be
	\begin{align}
	\nonumber \omega_{\text{peak}}&\simeq 10^3~\text{Hz}\left(\frac{\lambda\kappa_{\text{H}}}{4}\right)^{\frac{1}{6}}\left(\frac{T_{0}/\Lambda_{\text{H}}}{0.4}\right)^{\frac{2}{3}}\left(\frac{106.75}{g_{*,\text{ann}}}\right)^{\frac{1}{12}}\\
	&\times\left(\frac{10^{15}~\text{GeV}}{f}\right)^{\frac{1}{6}}\left(\frac{\Lambda_{\text{H}}}{10^{10}~\text{GeV}}\right),
	\end{align}
	while the peak amplitude of the gravitational wave signal is given by
	\begin{align}
	\Omega_{\text{gw}}h^2\simeq 10^{-17}\tilde{\epsilon}\left(\frac{\lambda}{4}\right)^{2}\left(\frac{106.75}{g_{*,\text{ann}}}\right)^{\frac{1}{3}}\left(\frac{f}{10^{15}~\text{GeV}}\right)^{4}.
	\end{align}
	From the formulae above, it appears to be possible to obtain a visible signal (at future interferometers) only when $f\lesssim 10^{16}~\text{GeV}$ and $\Lambda_{\text{H}}< 10^{10}~\text{GeV}$. In fact, there are two problems with these choices: First, such large values of $f$ may generally lead to PQ symmetry being never restored after inflation, since the current upper bound on the reheating temperature after inflation is $T\lesssim 10^{16}~\text{GeV}$. Secondly, reducing $\Lambda_{\text{H}}$ reduces the axion mass, which can then lead to a slower decay rate to SM states and an intermediate phase of MD, which has not been taken into account in deriving the equations above. When relevant, this suppresses the gravitational wave signal further. Overall, when including these effects, we find no observable region of parameter space for $N_{\text{DW}}=1$.
	
\end{document}